\documentclass[aps,prb,twocolumn,groupedaddress,showpacs]{revtex4}
\usepackage{epsfig}
\usepackage{amsmath}
\begin{document}

\title{Metallic phase in the two-dimensional ionic Hubbard model}
\author{K. Bouadim$^1$, N. Paris$^2$, F. H\'ebert$^1$,
G.G. Batrouni$^1$, and R.T. Scalettar$^2$}

\affiliation{$^1$INLN, Universit\'e de Nice Sophia--Antipolis, CNRS;
1361 route des Lucioles, 06560 Valbonne, France}

\affiliation{$^2$Physics Department, University of California, Davis,
California 95616, USA}

\begin{abstract}
We investigate the phases of the ionic Hubbard model in a
two-dimensional square lattice using determinant quantum Monte Carlo
(DQMC). At half-filling, when the interaction strength or the
staggered potential dominate we find Mott and band insulators,
respectively.  When these two energies are of the same order we find a
metallic region.  Charge and magnetic structure factors demonstrate
the presence of antiferromagnetism only in the Mott region, although
the externally imposed density modulation is present everywhere in the
phase diagram. Away from half-filling, other insulating phases are
found. Kinetic energy correlations do not give clear signals for the
existence of a bond-ordered phase.
\end{abstract}

\maketitle

\section{Introduction}

Tight binding models, such as the Hubbard Hamiltonian, have been
widely studied for their ability to describe different kinds of
insulating phases present in condensed matter systems. Interactions
between fermions give rise to gapped Mott insulators where the
fermions are immobile at commensurate fillings.  For bipartite
lattices, Mott behavior is typically associated with antiferromagnetic
order.  On the other hand, the imposition of an external periodic
potential can drive band insulating behavior. In systems where several
types of insulating effects are present, phase transitions can take
place as the relative strength of the parameters is changed.  Close to
such transitions, the competition between these different terms allows
smaller effects, such as kinetic energies, to result in low
temperature metallic phases or more exotic types of correlations such
as bond-order.  Such compensating effects are believed to explain
enhanced response in different quasi-one-dimensional systems such as
organic materials \cite{intro1}, ferroelectric perovskites
\cite{intro2} or transition metal-oxides \cite{intro3}.

A simple model exhibiting this kind of behavior is the ionic Hubbard
Hamiltonian \cite{hubbard81}, a tight binding model that includes an
on-site repulsion and a staggered potential that takes alternating
values on neighboring sites of a bipartite lattice. At half-filling,
the on-site repulsion drives the system into an antiferromagnetic Mott
insulator (MI), with a homogeneous density.  When the staggered
potential dominates, a band insulator (BI) with a regular modulation
of the density results.  A quantum phase transition between these two
states occurs in the phase diagram.

In one dimension, the ionic Hubbard model has been widely studied
through a variety of techniques including density matrix
renormalization group (DMRG), exact diagonalization, bosonization and
quantum Monte Carlo (QMC) with on-site
\cite{1dchain,zhang03,kampf03,manmana04} or extended
\cite{1dchainxtend,jeckelmann02,sengupta02,intro2,intro3,craig07}
interactions.  The precise conclusions are still subject to some
debate, but many of the studies suggest that the direct transition
from the band insulator to the Mott insulator is replaced by an
intervening insulating bond-ordered phase.  In two dimensions,
dynamical mean field theory (DMFT)\cite{garg06},
cluster-DMFT\cite{kancharla07} and determinant quantum Monte Carlo
(DQMC) \cite{paris07} were used.  The existence of an intermediate
phase between the BI and MI phases has been established, as in one
dimension, but the precise nature of the phase is still under debate
(see section II).  Studies of a bilayer Hubbard model
\cite{kancharla071} and in infinite dimension \cite{craco07} yield
similar results with a low temperature metal as the intermediate
phase.  Here the interplanar coupling, if sufficiently large, gives a
noninteracting band structure with a gap at half-filling much like a
staggered potential.  A recent experiment attempts to observe this
kind of phase \cite{maiti001}.

A related problem is the study of heterostructures of SrTiO$_3$ and
LaTiO$_3$, one being a MI, the other a BI and a metallic phase
appearing at the interface between the two \cite{hetero}.  Some of the
phenomena presented here also have similarities with those found in
corresponding disordered models used to describe binary alloys
\cite{binalloy}.

In this paper, we consider the different phases and transitions of the
ionic Hubbard Hamiltonian on a two-dimensional square lattice using
determinant quantum Monte Carlo (DQMC), extending our initial
study\cite{paris07}. In section II, we introduce the model and
summarize our previous results. In section III, we briefly describe
the numerical techniques used while in section IV, we investigate
further the different phases by measuring the charge and magnetic
structure factors and the spectral function.  We then present an
analysis of the properties away from half filling (section V) and the
search for bond-order waves (section VI). Conclusions are in section
VII.

\section{Fermions in a  Staggered potential}

We consider the ionic Hubbard model

\begin{align}
\hat H=&-t\sum_{\langle \mathbf{ r r'}\rangle \sigma} (c_{{\bf
r}\sigma}^\dagger c_{{\bf r'} \sigma} + c_{{\bf r'}\sigma}^\dagger
c_{{\bf r}\sigma}) +U \sum_{\bf r} n_{{\bf r}\uparrow} n_{{\bf
r}\downarrow} \nonumber \\ &+ \sum_{\bf r}
\left[\frac{\Delta}{2}(-1)^{(x+y)} - \mu \right] (n_{{\bf r} \uparrow}
+ n_{{\bf r} \downarrow})
\label{Hamiltonian}
\end{align}
\noindent
where $c_{{\bf r} \sigma}^{\dagger} (c_{{\bf r} \sigma})$ are the
usual fermion creation (destruction) operators on site ${\bf r} =
(x,y)$ of a two-dimensional square lattice with spin $\sigma$ and
where $n_{{\bf r} \sigma} = c_{{\bf r} \sigma}^{\dagger} c_{{\bf r}
\sigma}$ is the number operator. The hopping parameter, $t$, sets the
energy scale and is fixed at a value of $1$, while tuning the chemical
potential $\mu$ gives the desired fermion filling.  The interaction
strength $U$ and staggered potential $(-1)^{(x+y)} \, \Delta/2$ are
the two competing terms of our system, the former encouraging single
occupancy of all sites and the latter double occupancy of the low
energy sublattice.\cite{error}

Considering the limiting cases $U\gg \Delta$ and $\Delta \gg U$ we
find two distinct insulating phases at half-filling (density
$\rho=1$).  For $\Delta=0$, the usual translationally invariant
Hubbard model arises and, at $T=0$, forms an antiferromagnetic Mott
insulator at half-filling for any value of $U$.  At strong coupling,
the gap is set by $U$, while the effective anti-ferromagnetic coupling
is set by $J = 4t2/U$.  These two energy scales are evident in the
magnetic properties where $T \sim U$ demarks the onset of moment
formation on individual sites and $T \sim J$, the onset of
antiferromagnetic order between sites.

\begin{figure}[!t]
\centerline{\epsfig{figure=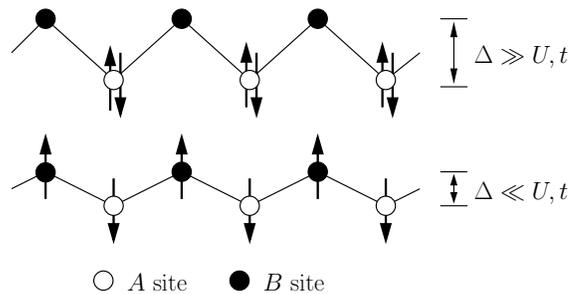,width=7.5cm,clip}}
\caption{Schematic diagram of electrons in a staggered potential.  The
$A$ sites have lower chemical potential than the $B$ sites (the
difference being $\Delta$). For low interaction strength $U\ll
\Delta$, the electrons prefer to doubly occupy the $A$ sites and form
a band insulator.  For $U\gg\Delta$, it is favorable to have only one
electron per site and to form a Mott insulator. The hopping term $t$
then generates an antiferromagnetic modulation.  }
\label{schem}
\end{figure}

In the non interacting limit $U=0$, the Hamiltonian can be
diagonalized in Fourier space (where it reduces to a set of
independent $2\times2$ matrices)
\begin{equation}
\nonumber
\hat{H} = \sum_{{\bf k},\sigma} E_{\bf k} \, c^\dagger_{{\bf
k}\sigma}c_{{\bf k}\sigma} \quad {\rm with} \quad E_{\bf k} = \pm
\sqrt{\epsilon_{\bf k}2+(\Delta/2)2}
\end{equation}
where the sum runs over half the Brillouin zone and where
$\epsilon_{\bf k} = -2t \left(\cos(k_x) + \cos(k_y)\right)$ is the
energy of a free particle of wave vector ${\bf k} = (k_x,k_y)$.  The
eigenenergies $E_{\bf k}$ are then split into two bands separated by a
gap $\Delta$ and, in this limit, the system is a band insulator at
half-filling.  These two insulating phases can be seen schematically
in Fig.~\ref{schem}.

\begin{figure}[!hb]
\centerline{\epsfig{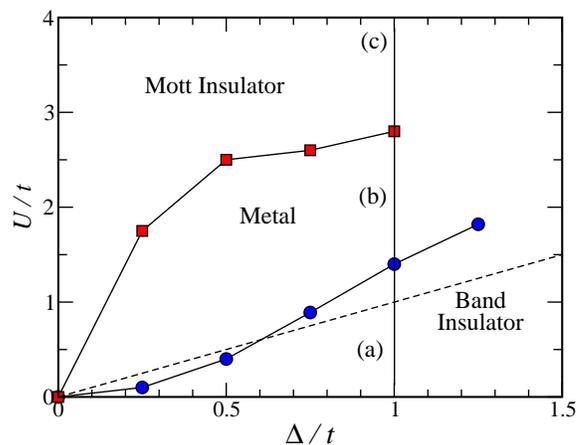}}
\caption{The phase diagram of the ionic Hubbard model\cite{error}.
Symbols are the result of QMC simulations. The dashed line is the
strong coupling ($t=0$) phase boundary between band and Mott
insulators.  Points (a), (b) and (c) located, respectively, in the BI,
the intermediate phase, and the MI are used in
Fig.~\ref{spectralfunc}.}\label{pd}
\end{figure}

Recently, it was shown\cite{garg06,kancharla07,paris07} that there is
no direct transition between these two insulating phases, but that an
intermediate phase appears (Fig. \ref{pd}). The precise nature of the
phase is still being debated.  In previous work \cite{paris07}, we
used the determinant quantum Monte Carlo algorithm (DQMC) to study the
conductivity of the model. Performing a finite size and temperature
analysis, we concluded that the intermediate phase has a finite
conductivity and is, consequently, metallic.  Garg {\it et al.} arrive
at the same conclusion in their dynamical mean field theory (DMFT)
study\cite{garg06} while using a cluster-DMFT analysis, Kancharla {\it
et al.} argue for a bond ordered wave (BOW) phase \cite{kancharla07}.

\begin{figure}[!ht]
\centerline{\epsfig{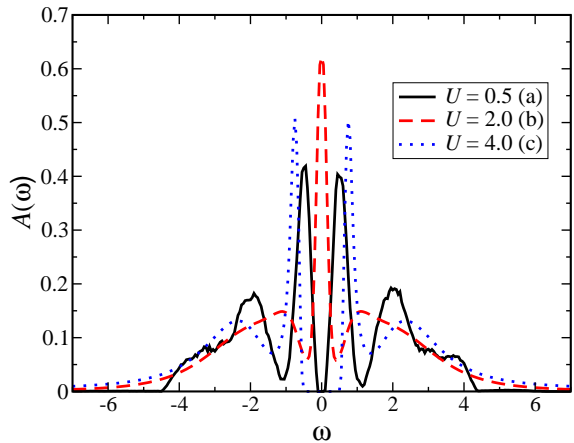}}
\caption{Spectral function $A(\omega)$ for various values of
interaction strength $U$ with $\Delta=1$ on a $6\times 6$ lattice at
$\beta=12$ at half-filling.  The letters (a)-(c) correspond to points
on the phase diagram Fig.~\ref{pd}. Point (b) exhibits a metallic
behavior as $A(\omega)$ is nonvanishing for $\omega=0$.  $A(\omega)$
exhibits clear gaps for points (a) and (c) which are, respectively,
band and Mott insulating states.}
\label{spectralfunc}
\end{figure}

\begin{figure}[!ht]
\centerline{\epsfig{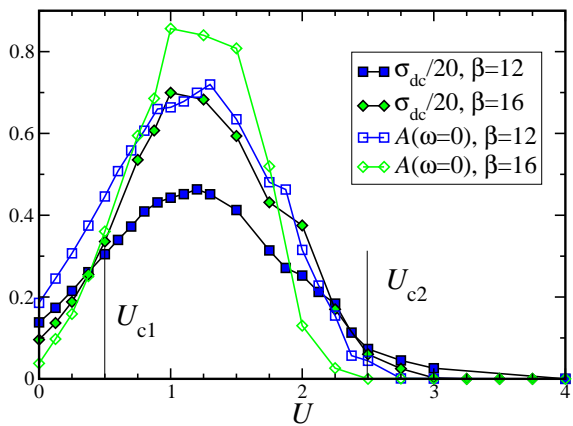}}
\caption{The spectral function $A(\omega)$ and conductivity
$\sigma_{\rm dc}$ are shown as functions of $U$ at $\beta=12$ and
$16$, at half-filling, and for $\Delta=0.5$. There is a BI-metal
transition at $U_{c1}\simeq0.5$ and a metal to MI transition at
$U_{c2}\simeq2.5$.  }
\label{specvsu}
\end{figure}

\section{Computational methods}

DQMC\cite{white89} is a numerical approach which expresses the
partition function $Z={\rm Tr}\, \exp(-\beta H)$ as a path integral
where the inverse temperature $\beta$ is discretized.  The interaction
term is decoupled via a Hubbard-Stratonovich (HS) transformation,
introducing a sum over an Ising spin field and leaving the Hamiltonian
in a quadratic form with respect to the fermion operators.  The
fermionic degrees of freedom can then be integrated out and the
remaining Boltzmann weight is expressed as the determinant of the
product of two matrices that depend on the HS spin variables.  In the
conventional Hubbard model, this determinant is positive at
half-filling, due to particle-hole symmetry.  With the staggered
potential, the determinant is no longer always positive at
half-filling resulting in a sign problem which can be severe.

The remaining sum over the HS field and the evaluation of observables
is achieved by a Metropolis Monte Carlo algorithm. It is difficult to
obtain good estimates for the observables when the determinant is
often negative, {\it i.e.}, when the mean value of the sign of the
determinant is small.  This problem is important for intermediate
values of $U$ and $\Delta$, while it is not severe in the limiting
cases where we retrieve simple BI or MI phases ($U$ or $\Delta$ small
or large).

With this formulation, space and (imaginary-)time separated Green
functions
\begin{equation}
\nonumber
G({\bf R},\tau) =  \left\langle c_{{\bf
r+R}\sigma}(\tau)\,c^\dagger_{{\bf r}\sigma}(0)\right\rangle
\end{equation}
are easily expressed as elements of the matrices.  >From the spatial
Fourier transform of $G({\bf R}, \tau)$, one can extract the spectral
function (essentially the density of states), by inverting the Laplace
transform
\begin{equation}
G({\bf k}=0, \tau) = \int d\omega \,
\frac{e^{-\omega\tau}}{1+e^{-\beta\omega}}A(\omega).
\label{GA}
\end{equation}
Solving for $A(\omega)$ remains a non-trivial task that is performed
with a method of analytic continuation\cite{anders98}.  We measure
correlation functions for charge and spin density waves (CDW and SDW,
respectively), given by
\begin{equation}
\nonumber C_{{\scriptstyle\rm cdw}\atop {\scriptstyle\rm sdw}}({\bf
R}) = \frac{1}{N}\sum_{\bf r} \left\langle \left[n_{{\bf r+R} \uparrow}
\pm n_{{\bf r+R} \downarrow}\right] \left[n_{{\bf r} \uparrow}\pm
n_{{\bf r} \downarrow}\right]\right\rangle\nonumber
\end{equation}
where $+$ corresponds to CDW and $-$ to SDW.  The staggered potential
imposes a checkerboard modulation on the density, not only in the BI
phase but in all phases as it breaks explicitly the translational
invariance.  Antiferromagnetic magnetic correlations are also expected
in the MI phase. Thus, for both types of correlations, we expect
ordering with a wave vector ${\bf k}_\pi = (\pi,\pi)$, characterized
by the divergence of corresponding structure factors
\begin{equation}
S_{{\scriptstyle\rm cdw}\atop{\scriptstyle\rm sdw}}= \sum_{\bf R} e^{i
{\bf k}_\pi\cdot{\bf R}} \, C_{{\scriptstyle\rm
    cdw}\atop{\scriptstyle\rm sdw}}({\bf R}). 
\end{equation}

\section{spectral function, magnetic and charge orders}

To complete our previous analysis\cite{paris07} of the intermediate
phase shown in Fig. \ref{pd} we evaluated the charge and spin
correlations and spectral function.  When the size is not explicitly
mentioned, we used a $N=6 \times 6 = 36$ sites lattice. $N_x =
\sqrt{N}$ denotes the side of the lattice.

Fig.~\ref{spectralfunc} shows the spectral function for $\Delta=1$
(along the vertical line shown in Fig. \ref{pd}) at $\beta=12$ and for
three values of $U$ corresponding to the three different phases.  For
$U=0.5$ and $U=4$, corresponding to the BI and MI phases, the spectral
function shows a gap around $\omega=0$, thus indicating an
insulator. In contrast, the spectral function for $U=2$ is non zero
and shows a quasi-particle peak at $\omega=0$ which indicates metallic
behavior.

\begin{figure}[!ht]
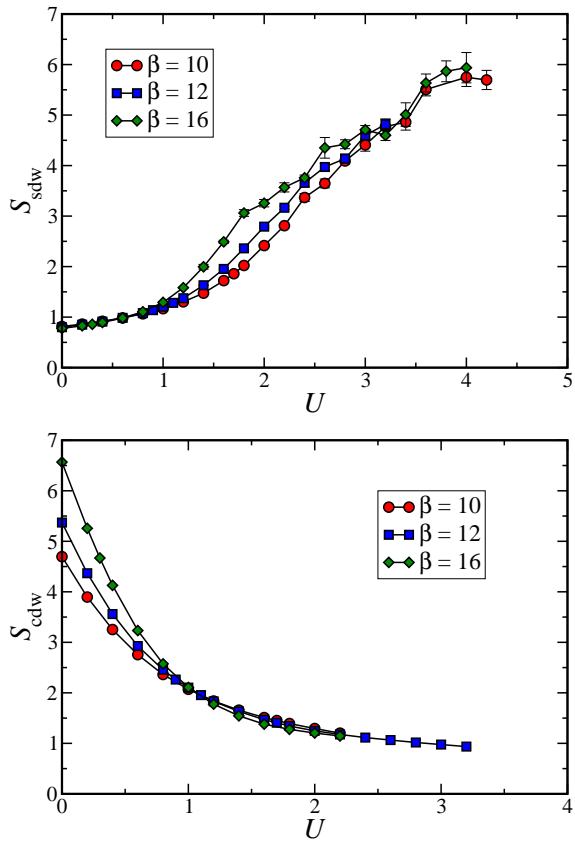

\centerline{\epsfig{figure=fig5a.eps,width=7.5cm,clip}}

\medskip 

\centerline{\epsfig{figure=fig5b.eps ,width=7.5cm,clip}}
\caption{Top: Antiferromagnetic structure factor $S_{\rm sdw}$ as a
function of repulsion $U$ shows antiferromagnetic order developing for
large $U$.  Bottom: Conversely, the charge structure factor $S_{\rm
cdw}$ decreases as $U$ increases.  $\Delta=0.25$ in both figures.}
\label{structfac}
\end{figure}

These results are consistent with measurements of the conductivity in
our previous article, as is confirmed by the direct comparison of the
spectral function at zero frequency $A(\omega=0)$ and the conductivity
$\sigma_{\rm dc}$ shown in Fig.~\ref{specvsu}.  For $\Delta=0.5$ and
as $U$ is increased, the behaviors of these two quantities are
similar: they both show, for intermediate values of $U$, a peak that
grows with $\beta$. The ranges of interaction where there are peaks
are slightly different because of finite size effects.

$U_{c1}$ and $U_{c2}$ are the critical repulsion energies that cause
BI to metal and metal to MI phase transitions for a fixed value of
$\Delta$.  These critical values are obtained by a finite temperature
analysis of the conductivity\cite{paris07}. As the temperature is
lowered, the conductivity increases in a metal and decreases in an
insulator. Locating the points where the behavior of the conductivity
changes gives the critical values.  For $\Delta=0.5$, the BI-metal
transition is located at $U_{c1}/t\simeq0.5$, the metal-MI transition
is located at $U_{c2}/t\simeq2.5$ and the metallic region is in
between.

\begin{figure}[!ht]
\centerline{\epsfig{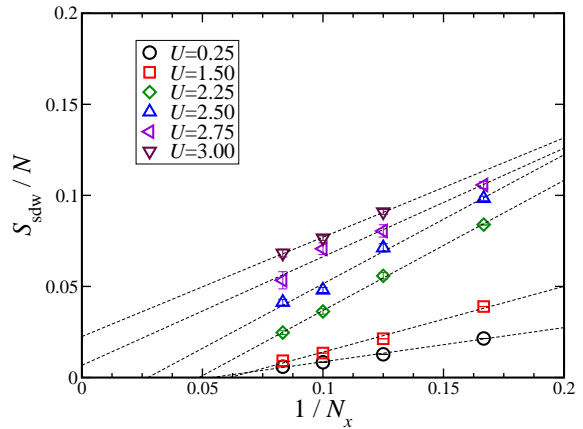}}
\caption{The antiferromagnetic structure factor $S_{\rm sdw}/N$ as a
function of inverse lattice size $1/N_x$ at $\Delta=0.5$ and
$\beta=12$ for various $U$.  The different values of $U$ correspond to
different phases: $U=0.25$ to the BI, $U=1.50, 2.25$ and $2.5$ to the
metal, and $U=2.75$ and $3$ to the MI.  The lines are least-squares
fit to the data and the extrapolation leads to the value of $S_{\rm
sdw}=m2/3$ for an infinite system. There is antiferromagnetic long
range order only in the MI phase.}
\label{finitesizeAF}
\end{figure}

Structure factors $S_{\rm sdw}$ and $S_{\rm cdw}$ are shown in
Fig.~\ref{structfac} for $\Delta$= 0.25 and for inverse temperatures
$\beta=10, 12$ and 16.  For these values of $\Delta$, the metallic
boundaries are $U_{c1}\simeq 0.1$ and $U_{c2} \simeq 1.7$. As $U$ is
increased, $S_{\rm sdw}$ increases as the system is driven towards the
MI phase, while $S_{\rm cdw}$ decreases, in moving away from the BI
phase.  The transitions in these quantities when we cross the phase
boundaries are not sharp on lattices of fixed, finite size owing to
finite size rounding effects.  In addition, as stated above, a
smoother evolution is expected for $S_{\rm cdw}$ as a modulation of
$\rho$ is imposed by the external potential for all parameter values.

In a conventional Hubbard model, the MI at half filling is always an
antiferromagnet.  However, it has been suggested recently for related
models \cite{kancharla071} that the metal-MI and para- to
antiferro-magnetic (PM-AF) transitions may not take place together.  A
finite size scaling analysis of $S_{\rm sdw}$ for different values of
$U$ is needed to check whether these two boundaries coincide in our
model.  The structure factor $S_{\rm sdw}$ increases like the number
of sites $N$ in the thermodynamic limit
\begin{align}
\frac{S_{\rm sdw}}{N} = \frac{m^{2}}{3} + O \left( \frac{1}{N_x}
\right)
\label{eq.structspin}
\end{align}
where $m2$ is the antiferromagnetic order parameter.
Fig.~\ref{finitesizeAF} shows structure factors for sizes up to
$N=12\times12$ sites for different values of $U$. $\Delta$ is set to
0.5 and the metallic phase appears between $U_{\rm c1}\simeq 0.5$ and
$U_{\rm c2}\simeq 2.5$.  We apply a least squares fit to the data, and
extrapolate to an infinite system.  We find a non-vanishing structure
factor only in the MI phase ($U=2.75$ and $3$). In all the other
cases, including points located close to the metal-MI transition
($U=2.25$ and $2.5$), the structure factor goes to zero as the size is
increased.

The PM-AF transition, therefore, takes place between $U=2.5$ and
$U=2.75$ and appears to be simultaneous with the metal-MI transition.
Although our study on limited sizes cannot completely exclude
completely the possibility of an exotic AF metallic phase located
around the metal-MI boundary, our data suggest that it does not occur.

\begin{figure}[!ht]
\centerline{\epsfig{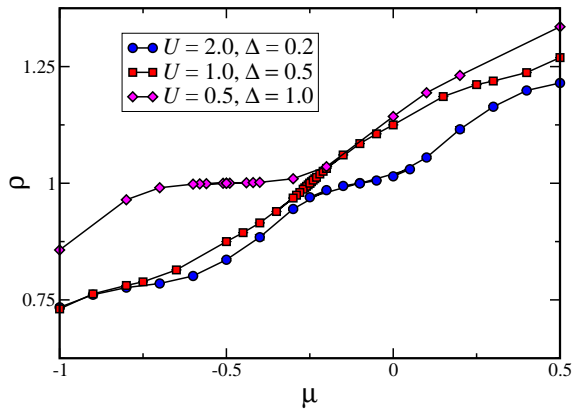}}
\caption{Density, $\rho$, as a function of chemical potential $\mu$
for the different phases.  The Mott ($U=2.0, \Delta=0.2$, circles) and
band ($U=0.5, \Delta=1.0$, diamonds) insulating phases have plateaux
as the compressibility $\kappa=\partial\rho/\partial\mu=0$ at
half-filling.  For the metallic phase ($U=1.0,\Delta=0.5$, squares),
there is no plateau in the region around $\rho=1$.}
\label{rhovmu1}
\end{figure}

\section{away from half-filling}

\begin{figure}[!ht]
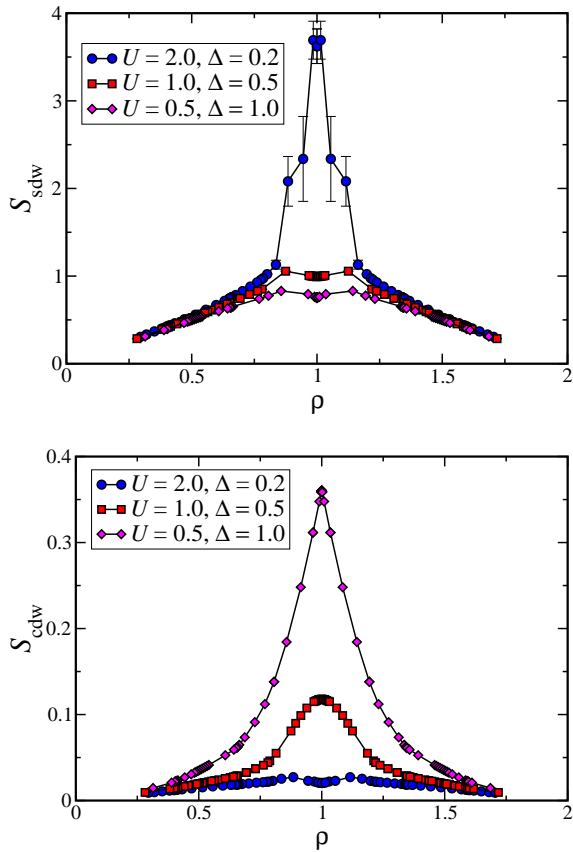

\centerline{\epsfig{figure= fig8a.eps,width=7.5cm,clip}}

\bigskip

\centerline{\epsfig{figure= fig8b.eps,width=7.5cm,clip}}
\caption{Spin (top) and charge (bottom) structure factors as functions
of density.  The parameters are the same as in Fig.~\ref{rhovmu1}.  As
$\rho$ approaches half-filling, $S_{\rm sdw}$ reaches its maximum
value for the Mott phase.  Conversely, $S_{\rm cdw}$ is maximum at
$\rho=1$ for the band insulator.  The metallic phase has a slight peak
in $S_{\rm cdw}$ but not in $S_{\rm sdw}$.}
\label{structvrho}
\end{figure}

We extend our study to determine the behavior away from half-filled
band.  In Fig.~\ref{rhovmu1}, the density $\rho$ is plotted as a
function of chemical potential. The compressibility $\kappa$ is given
by $\partial\rho/\partial\mu$.  For $\kappa$=0, incompressible
insulating phases are present which is indicated by plateaux in the
$\rho$ versus $\mu$ curve.  Both the MI ($U=2.0,\Delta=0.2$) and BI
($U=0.5,\Delta=1.0$) phases have this distinctive plateau at
half-filling.  The metallic nature of the intermediate phase is once
again confirmed by the absence of a plateau around $\rho=1$.

Using the same parameters $U$ and $\Delta$, we show in
Fig.~\ref{structvrho} the evolution of the structure factors as
functions of $\rho$.  As expected, for the parameters corresponding to
the MI at $\rho=1$, $S_{\rm sdw}$ shows a strong peak around
half-filling indicating the presence of the AF-MI in this case and its
rapid suppression upon doping. In this case, charge correlations
always remain small, as double occupancy of sites is essentially
forbidden.  Similarly, for the parameters corresponding to the BI, the
peak is present in $S_{\rm cdw}$ at $\rho=1$ and $S_{\rm sdw}$ remains
small for all densities, as the occupation of larger energy site is
suppressed.  For the metallic case, there is no peak in the magnetic
structure factor at $\rho=1$ but a moderate peak in $S_{\rm cdw}$.

In all these three cases, away from half-filling, the structure
factors are rapidly suppressed and the system is compressible and
becomes a paramagnetic metal.

\begin{figure}[!ht]
\centerline{\epsfig{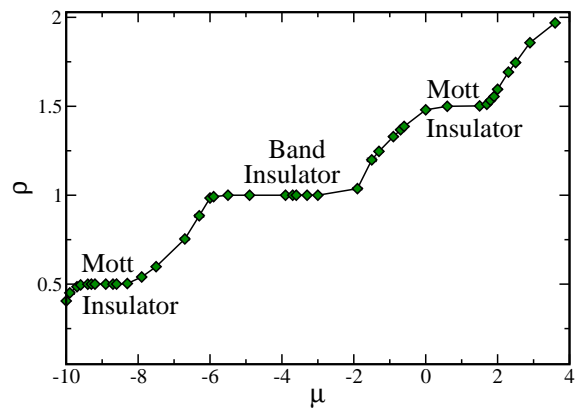}}
\caption{The density, $\rho$, as a function of chemical potential
$\mu$ for $\Delta=8$ and $U=4$ at $\beta=10$.  The plateaux represent
the various insulating phases.  The regions where $\kappa>0$ are
metallic phases.}
\label{rhovmu2}
\end{figure}

\begin{figure}[!ht]
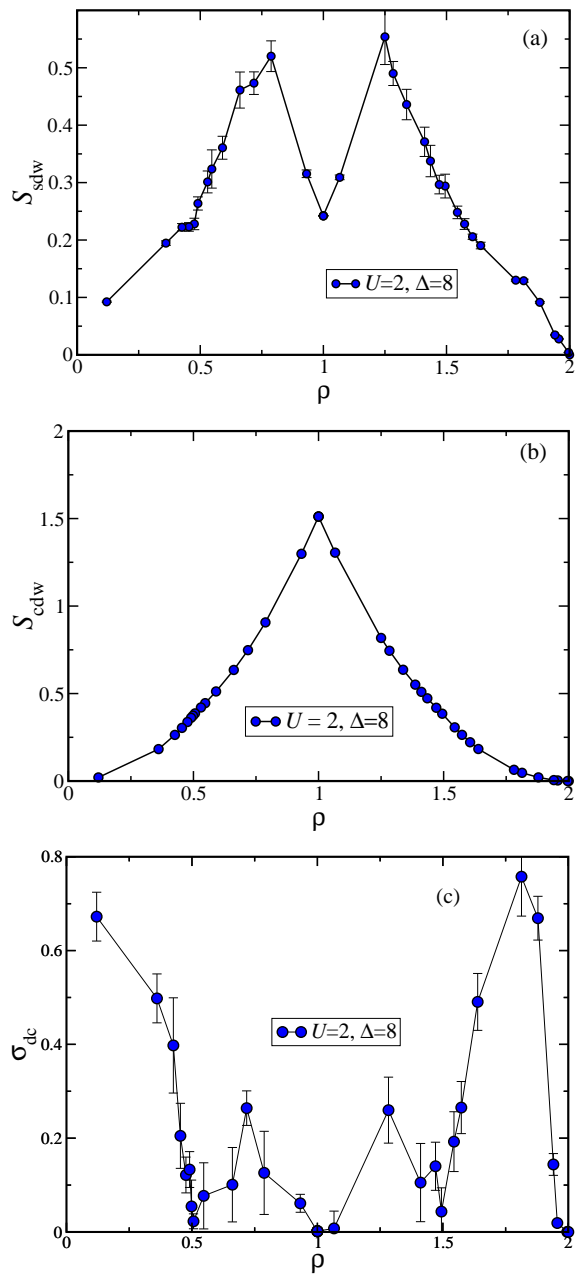

\centerline{\epsfig{figure= fig10a.eps,width=7.5cm,clip}}

\medskip

\centerline{\epsfig{figure= fig10b.eps,width=7.5cm,clip}}

\medskip

\centerline{\epsfig{figure= fig10c.eps,width=7.5cm,clip}}
\caption{(a) Antiferromagnetic structure factor $S_{\rm sdw}$, (b)
charge structure factor $S_{\rm cdw}$ and (c) conductivity,
$\sigma_{\rm dc}$ as functions of $\rho$ at $U=2$, $\Delta=8$ and
$\beta=10$. Note that $\sigma_{\rm dc}$ vanishes in the three solid
phases, $\rho=1/2,\, 1,\,3/2$.}
\label{structvmu}
\end{figure}

For stronger interactions and staggered potential, other ordered
phases can be found at quarter and three quarters fillings.  For $U=4$
and $\Delta=8$, there is a band insulator at half-filling shown by the
$\rho=1$ plateau in Fig.~\ref{rhovmu2}.  Additional plateaux emerge at
$\rho =0.5$ and $1.5$.  The plateau at $\rho=0.5$ corresponds to a
state where all the lower energy sites are occupied by one fermion.
Adding a fermion to an already occupied site increases the energy
approximately by $U$, while adding it to an empty (high energy) site
increases it by $\Delta$. There is then an energy gap, roughly equal
to the smaller of the two parameters ($U$ in this case).  We then have
a Mott insulator at quarter filling and, symetrically, another one at
three quarters filling. In the case where $U$ is larger than $\Delta$,
the situation is reversed and we expect a MI at half-filling and a BI
for $\rho=0.5$ and $\rho=1.5$ (this case is difficult to treat
numerically because of a strong sign problem).

Although the system is incompressible for $\rho=0.5$, there is no
concommitant magnetic order: in Fig.~\ref{structvmu}, $S_{\rm sdw}$
always remains small as $\rho$ is varied away from half-filling and
there is no AF peak around quarter filling (the maximum being around
$\rho=0.75$). The AF ordering appears when the fermions jump between
neighboring sites, thus creating an effective antiferromagnetic
coupling between spins. At quarter filling, the particles are
surrounded by empty sites and there is no effective coupling. As for
$S_{\rm cdw}$ we only observe a peak at half-filling
(Fig.~\ref{structvmu}). A plot of the conductivity
(Fig.~\ref{structvmu}(c)) versus $\rho$ shows that, indeed,
$\sigma_{\rm dc}$ vanishes for $\rho=1/4$ and $\rho=3/4$ corresponding
to the plateaux observed in Fig.~\ref{rhovmu2}.

\section{bond-ordered waves}

Long range Bond-ordered waves (BOW) are known to exist in one
dimension in the ionic Hubbard model \cite{manmana04,kampf03,zhang03}
as well as in other one-dimensional systems exhibiting density
modulation such as extended Hubbard models
\cite{jeckelmann02,sengupta02}.  (In the latter case, the two
numerical treatments, using the stochastic series
expansion\cite{sengupta02} and density matrix renormalization
group\cite{jeckelmann02}, disagree on the precise width of the BOW
phase as well as over what range of coupling values it is present.)
It was suggested by Kancharla {\it et al.}\cite{kancharla07} that the
intermediate phase observed at half-filling in the two-dimensional
case studied here was in fact a bond-ordered state, as in one
dimension.  Other studies have attempted to observe bond-order in two
dimensions, for example in extended Hubbard models
\cite{murakami00}. In two dimensions, bond orders have mostly been
obtained in Hamiltonians possessing rather special geometries or
interactions, e.g.~on checkerboard lattices \cite{indergand07} or by
including additional terms such as ring exchanges \cite{melko04}.

\begin{figure}[!ht]
\centerline{\epsfig{figure= 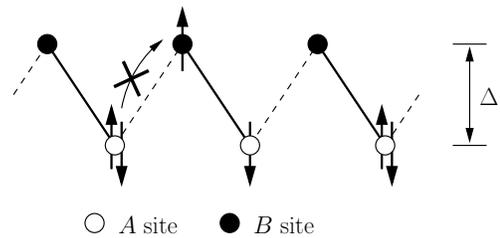,width=6.5cm,clip}}
\caption{Schematic diagram of how a bond-ordered wave occurs in one
dimension.  For $U\sim \Delta$ both large, the electrons may move to a
site of higher chemical potential as shown with the spin-up electron.
The spin-down electron in the left well is not likely to move to the
right because it will cost a large energy $\Delta$ to doubly occupy
the site of higher chemical potential.  Prohibited movement to left
causes a decrease in kinetic energy for that bond. This alternation of
two bonds with weak and strong kinetic energy bonds is an indication
of the appearance of a bond order.  Strong bonds are continuous thick
lines, weak bonds are dotted lines.  }
\label{bowschem}
\end{figure}

\begin{figure}[!t]
\centerline{\epsfig{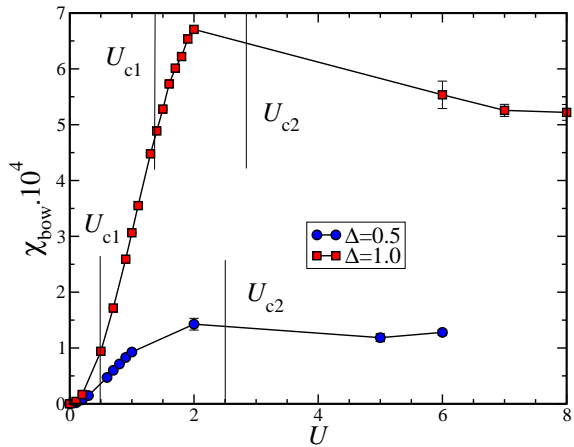}}
\caption{The bond-ordered susceptibility as a function of $U$ at
$\rho=1$ for $N=12\times 12$ and $\beta=10$.  The vertical lines
separate the three different phases for two different values of the
staggered potential.  The maximum of this bond-order parameter is in
the metallic region but is of order $10^{-4}$.  } \label{corrbow}
\end{figure}

\begin{figure}[!t]
\centerline{\epsfig{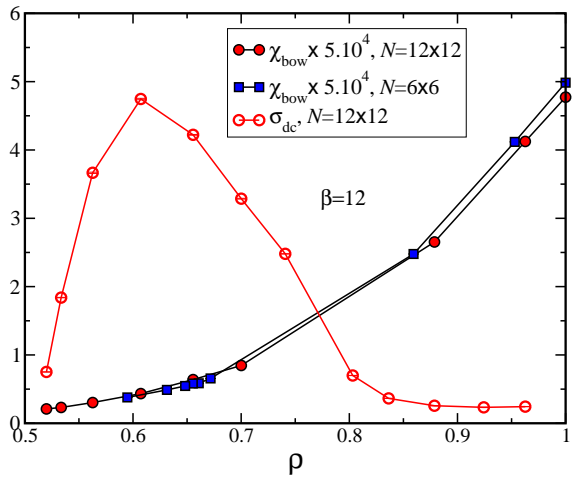}}
\caption{The bond-ordered susceptibility and conductivity as functions
  of $\rho$ for $\Delta=1$, $U=0.5$, and $\beta=12$. Doping away from
  half-filling, the susceptibility is non-zero even when the band is
  only partially full and the system is unquestionable metallic.}
\label{corrbow2}
\end{figure}

Hence, bond order seems in general more difficult to obtain in two
dimensions than in one. In our system, we can understand this with the
argument sketched in Fig.~\ref{bowschem}.  Starting from a situation
where all the particles are located on the low energy sites ($A$
sites), a jump to a high energy site ($B$ site) increases the energy
roughly by $\Delta - U$.  If $\Delta$ and $U$ are of the same order of
magnitude, this energy cost is small and the jump has a high
probability to happen.  This $B$ site is connected to two $A$ sites:
the one where the jumping particle comes from (to the right in
Fig.~\ref{bowschem}) and another one (to the left) still occupied by
two particles. The jumps on the left bond are suppressed: the particle
on the $B$ site and the particle having the same spin (an up spin in
the sketch) on the $A$ site cannot jump because of Pauli principle,
while a jump of the remaining particle has a large energy cost
$\Delta$.

In short, having a particle jumping on a bond (and thus a contribution
to the kinetic energy on this bond) will inhibit jumps on the
following bond. In one dimension, this can lead to a bond order with
alternate strong and weak values of the kinetic energy. On a two
dimensional square lattice, the appearance of a strong bond will
inhibit the three other bonds linked to the same site and it seems
more difficult to generate regular arrays of weak and strong bonds.

To check if there is a BOW, we examined the correlation function of
the kinetic energy
\begin{equation}
\nonumber
K^{xx}({\bf R}) = \sum_{\bf r}\langle k^{x}_{\bf
r+R}k^{x}_{\bf r} \rangle, 
\end{equation}
where $k^{x}_{\bf r}$ is the kinetic energy operator connecting sites
${\bf r}$ and ${\bf r}+\hat{x}$ ($\hat x$ being a unit vector directed
along the $x$ direction),
\begin{equation}
\nonumber
k_{\bf r}^x = \sum_{\sigma =\uparrow, \downarrow} (c_{{\bf
r}+\hat{x},\sigma}^{\dagger}c_{{\bf r},\sigma} + c_{{\bf
r},\sigma}^{\dagger}c_{{\bf r}+\hat{x},\sigma}).
\end{equation}
The Fourier transform of $K^{xx}({\bf R})$ always exhibits a very
small peak for ${\bf k}_\pi=(\pi,\pi)$, in {\it } all the different
phases at or away from half-filling. Finite size scaling analysis
indicates that these peaks persist in large size lattices, but this is
the case for all the different phases, not only in the intermediate
phase.  We also studied the BOW susceptibility\cite{sengupta02} for
this wave vector
\begin{equation}
\chi_{\rm bow} = \sum_{{\bf R,r},\tau}\, e^{i\, {\bf k}_\pi\cdot{\bf
R}} \left\langle k_{\bf r+R}^x(\tau) k_{\bf r}^x(0) \right\rangle
\end{equation}
Fig.~\ref{corrbow} shows the evolution of $\chi_{\rm bow}$ at
half-filling as a function of $U$ and for two different values of
$\Delta$.  The curve is maximum in the intermediate phase but is also
non zero in other phases. Similar results are obtained for bonds
oriented along the $y$ direction and for mixed $x$-$y$ cases.

We do not believe these signals to indicate the presence of a BOW
phase for the following reasons (a) The values obtained are extremely
small. (b) These ``BOW signals'' are ubiquitous in {\it all three
phases} and not specific to the intermediate one. The susceptibility
is, for example, nearly as important in the MI phase as in the
intermediate phase.  As soon as $U$ and $\Delta$ are both non-zero, we
observe this very weak non-zero value of $\chi_{\rm bow}$. (c)
Starting from the metallic phase ($U=0.5, \Delta=1$) and doping away
from half-filling the system is certainly known to be metallic (see
Fig.~\ref{corrbow2}) yet the BOW susceptibility is non zero and
comparable to the value found at half-filling.  For all these reasons,
we believe that these tiny signals are only a weak signature of an
interplay between the interaction and externally imposed checkerboard
potential.

\section{conclusions}

We have extended our previous study of the phase diagram of the ionic
Hubbard model \cite{paris07}. In addition to new QMC results for the
conductivity, we calculated the spectral function, the kinetic energy
correlation function (the order parameter for the BOW phase) and the
magnetic and charge order parameters. Our results point to the
presence of a normal conducting metallic phase intervening between the
band and Mott insulating phases. In particular, we have not found a
clear signal in the BOW order parameter and susceptibilities
indicating the presence of such a phase.

We also studied the system doped away from half filling. When the band
or Mott insulating phase is doped, a conducting phase is obtained
except at fillings of $1/4$ and $3/4$ where we found insulating phases
without magnetic order.

G.G.B., F.H., and K.B. are supported by a grant from the CNRS (France)
PICS 18796, R.T.S. and N.P. were supported by NSF ITR 0313390.  We
acknowledge helpful discussions with M.M. Gibbard.


\begin{thebibliography}{10}



\bibitem{intro1}
N.~Nagaosa and J.~Takimoto, 
J. Phys. Soc. Jpn {\bf 55}, 2735 (1986) ;
{\it ibid.} {\bf 55}, 2745 (1986) ; {\it ibid.} {\bf 55}, 2745 (1986) and references therein.

\bibitem{intro2}
T.~Egami, S.~Ishihara and M.~Tachiki, 
Science {\bf 261}, 130 (1994).

\bibitem{intro3}
S.~Ishihara, T.~Egami and M.~Tachiki, 
Phys.~Rev.~B {\bf 49}, 8944 (1994).



\bibitem{hubbard81}
J.~Hubbard and J.B.~Torrance,
Phys.~Rev.~Lett. {\bf 47}, 1750 (1981).

 
\bibitem{1dchain}
T.~Wilkens and R.M.~Martin,
Phys.~Rev.~B {\bf 63}, 235108 (2001);
M.E.~Torio, A.A.~Aligia, and H.A.~Ceccatto,
Phys.~Rev.~B {\bf 64}, 121105(R) (2001);
C.D.~Batista and A.A.~Aligia,
Phys.~Rev.~Lett. {\bf 92}, 246405 (2004);
G.I.~Japaridze, R.~Hayn, P.~Lombardo, and E.~M\"{u}ller-Hartmann,
cond-mat/0611415 (2006).

\bibitem{kampf03}
A. P. Kampf, M. Sekania, G. I. Japaridze, and Ph. Brune,
J. Phys.: Condens. Matter {\bf 15} 5895-5907 (2003).

\bibitem{zhang03}Y.Z.~Zhang, C.Q.~Wu, and H.Q.~Lin,
Phys.~Rev.~B {\bf 67}, 205109 (2003).

\bibitem{manmana04}
S.R.~Manmana, V.~Meden, R.M.~Noack, and K.~Sch\"{o}hammer,
Phys.~Rev.~B {\bf 70}, 155115 (2004).


\bibitem{1dchainxtend}
A.A.~Aligia and C.D.~Batista,
Phys.~Rev.~B {\bf 71}, 125110 (2005).

\bibitem{jeckelmann02}
E.~Jeckelmann,
Phys.~Rev.~Lett. {\bf 89}, 236401 (2002).

\bibitem{sengupta02}
P.~Sengupta, A.W.~Sandvik, and D.K.~Campbell,
Phys.~Rev.~B {\bf 65}, 155113 (2002).

\bibitem{craig07}
H. Craig, C.N. Varney, W.E. Pickett, and R.T. Scalettar,
work in progress.

\bibitem{garg06}
A.~Garg, H.R.~Krishnamurthy, and M.~Randeria,
Phys.~Rev.~Lett {\bf 97} 046403 (2006).

\bibitem{kancharla07}
S.S.~Kancharla and E.~Dagotto,
Phys.~Rev.~Lett. {\bf 98}, 016402 (2007).

\bibitem{paris07}
N.~Paris, K.~Bouadim, F.~H\'ebert, G.G.~Batrouni, and R.T.~Scalettar,
Phys.~Rev.~Lett. {\bf 98}, 046403 (2007).

\bibitem{error}
Our previous article \cite{paris07} contained a factor
of two error in the value of the staggered potential $\Delta$.
The correct definition of the Hamiltonian is the one given here. 
The BI-MI transition line in the $t=0$ limit
in Fig.~\ref{pd} has been modified accordingly.


\bibitem{kancharla071}
S.S.~Kancharla and S. Okamoto, cond-mat/0703728.

\bibitem{craco07}
L.~Craco, P.~Lombardo, R.~Hayn, G.I.~Japaridze, and E.~M\"uller-Hartmann,
cond-mat/0703814.

\bibitem{maiti001}
K.~Maitim, R.~Shankar Singh, and V.R.R.~Medicherla, cond-mat/0704.0327.

\bibitem{hetero}
S.S.~Kancharla and E.~Dagotto,
Phys.~Rev.~B {\bf 74}, 195427 (2006);
S.~Okamoto and A.J.~Millis,
Phys.~Rev.~B {\bf 70}, 075101 (2004).


\bibitem{binalloy}
K.~Byczuk, W.~Hofstetter, and D.~Vollhardt,
Phys.~Rev.~B {\bf 69}, 045112 (2004);
P.~Lombardo, R.~Hayn, and G.I.~Japaridze,
Phys.~Rev.~B {\bf 74}, 085116 (2006);
N.~Paris, A.~Baldwin, and R.T.~Scalettar,
cond-mat/0607427 (2006).


\bibitem{white89}
S.R.~White, D.J.~Scalapino, R.L.~Sugar,E.Y.~Loh, J.E.~Gubernatis, and R.T.~Scalettar,
Phys.~Rev.~B {\bf 40}, 506 (1989).

\bibitem{anders98}
A.W.~Sandvik, Phys.~Rev.~B {\bf 57}, 10287 (1998).


\bibitem{murakami00}
M.~Murakami,
Journal of the Physical Society of Japan {\bf 69}, 1113 (2000)

\bibitem{indergand07}
M.~Indergand, C.~Honerkamp, A.~La\"uchli, D.~Poilblanc, 
and M.~Sigrist, Phys. Rev. B {\bf 75}, 045105 (2007).

\bibitem{melko04}
R.G.~Melko, A.W.~Sandvik, and D.J.~Scalapino, 
Phys. Rev. B {\bf 69}, 100408 (2004).



\end{thebibliography}
\end{document}